\documentclass[a4paper,fleqn,usenatbib]{mnras}
\usepackage[T1]{fontenc}
\usepackage{ae,aecompl}
\usepackage{graphicx}
\usepackage{amsmath}
\usepackage{amssymb}
\usepackage{here}
\usepackage{color}
\title[Radio broadband visualization of simulated spiral galaxies I]{Radio broadband visualization of global three-dimensional magneto-hydrodynamical simulations of spiral galaxies I. Faraday rotation at 8~GHz
}
\author[M. Machida et al.]{
M. Machida,$^{1}$ 
T. Akahori,$^{2}$
K. E. Nakamura,$^{4}$ 
H. Nakanishi,$^{2,5,6}$  
and M. Haverkorn $^{7}$ 
\\
$^{1}$Department of Physics, Faculty of Sciences, Kyushu University,  744 Motooka, Nishi-ku, Fukuoka, 819-0395, Japan 
\thanks{mami@phys.kyushu-u.ac.jp} \\
$^{2}$Graduate School of Science and Engineering, Kagoshima University, Korimoto 1-21-35, Kagoshima 890-0065, Japan\\
$^{3}$National Astronomical Observatory of Japan, 2-21-1 Osawa, Mitaka, Tokyo 181-8588, Japan \\
$^{4}$Kyushu Sangyo University, 3-1 Matsukadai 2-chome, Higashi-ku, Fukuoka, 813-8503, Japan\\
$^{5}$Institute of Space and Astronautical Science, Japan Aerospace Exploring Agency, 3-1-1 Yoshinodai, Sagamihara, Kanagawa 252-5210, Japan\\
$^{6}$SKA Organization, Jodrell Bank Observatory, Lower Withington, Macclesfield, Cheshire SK11 9DL, UK\\
$^{7}$1Department of Astrophysics / IMAPP, Radboud University Nijmegen, PO Box 9010, 6500 GL Nijmegen, The Netherlands \\
}

\date{Accepted XXX. Received YYY; in original form ZZZ}
\pubyear{2016}
\begin{document}
\label{firstpage}
\pagerange{\pageref{firstpage}--\pageref{lastpage}}
\maketitle

\begin{abstract}

Observational study of galactic magnetic fields is hampered by the fact that the observables only probe various projections of the magnetic fields.
Comparison with numerical simulations is helpful to understand the real structures, and observational visualization of numerical data is an important task. In this paper, we investigate 8~GHz radio synchrotron emission from spiral galaxies, using the data of global three-dimensional magneto-hydrodynamic simulations. 
We assume a frequency independent depolarization in our observational visualization. 
We find that the appearance of the global magnetic field depends on the viewing angle: a face-on view seemingly has hybrid magnetic field types combining axisymmetric modes with higer order modes; at a viewing angle of $\sim 70\degr$, the galaxy seems to contain a ring-like magnetic field structure; while in edge-on view, only field structure parallel to the disk can be seen. 
The magnetic vector seen at 8~GHz traces the global magnetic field inside the disk. 
These results indicate that the topology of global magnetic field obtained from the relation between azimuthal angle and Faraday depth strongly depends on the viewing anglue of the galaxy. As one of the examples,  
we compare our results at a viewing angle of $25\degr$ with the results of IC342. 
The relation between azimuthal angle and Faraday depth of the numerical result shows a tendency similar to IC342, such as the peak numbers of the Faraday depth. 
\end{abstract}


\section{Introduction}

The origin and evolution of magnetic fields in spiral galaxies is a longstanding question. Radio observations of synchrotron radiation and Faraday rotation have provided various pieces of information of magnetic fields perpendicular to and parallel to the line of sight (LOS), respectively. The global magnetic field of spiral galaxies has been classified into axi-symmetric spiral (ASS) seen in M31 or bi-symmetric spiral (BSS) in M81, based on the topology. Since the discovery of hybrid (ASS and BSS) structure in M51, many galaxies are classified into hybrid-types which consist of ASS and higher modes \citep{fle2011, bec2016}. Such global magnetic fields are also known to co-exist with turbulent magnetic fields
with comparable magnetic energy densities, which correspond to field strengths of microgauss.
(e.g., \citealt{bec2007}). 

Such galactic magnetic fields are produced for billions of years by the maintenance mechanisms which are related to the field topology. \citet{par1970} and \citet{par1971} proposed a galactic dynamo known as the alpha-omega dynamo. That mechanism is able to produce toroidal magnetic fields from poloidal fields by the $\alpha$-effect and poroidal fields from toroidal fields by the $\Omega$-effect.  It can amplify weak seed field into microgauss field over $10^8$ years 
converting kinetic energy of the differentially rotating disk and energy from associated turbulent gas motions into magnetic energy.  Extended dynamo theories were proposed in the 1980s.  These models explain both ASS and BSS features of magnetic fields \citep{chi92, fuj1987}.

More recently, the importance of the magneto-rotational instability (MRI) was recognized \citep{bal1991}. Its effect on gaseous motions is taken into account in several global simulations of a galactic gas disk \citep{nis2006, han2009, mac2009, mac2013}. For instance, \citet{mac2013} performed a three-dimensional (3D) magneto-hydrodynamic (MHD) simulation of the galactic gas disk. They demonstrated the evolution of the galactic disk induced by the dynamo effect according to the MRI and Parker instabilities. \citet{gre2013} carried out hybrid dynamo simulations in which they considered the effects of both the MRI and the $\alpha-\Omega$ dynamo excited by supernova explosions. Recently, the formation and evolution of a Milky Way-like disk galaxy is researched in full cosmological context with magnetic fields \citep{pak2014}. 

In order to examine the topology of magnetic fields and understand their origin and evolution, theoretical predictions of observables are important. For example, using data of 3D MHD turbulence simulations, high-precision modeling and observational visualization of diffuse ionized gas toward high Galactic latitudes was conducted \citep{aka2013}. Wide-field and high-spatial-resolution radio observations with future radio interferometers such as the Square Kilometre Array (SKA) would dramatically improve imaging quality in galaxy surveys, and would allow to study complex structures that reflect the MRI-Parker instability. Moreover, future radio observations will achieve wideband polarimetry, which will deliver a new research dimension and will become a breakthrough of the study of galactic magnetic fields (e.g., \citealt{heald2015}). For example, the use of Faraday tomography for astronomical objects has started (see e.g., \citealt{aka2014,ide2014}). 

Radio broadband visualization is, however, challenging because the effect of depolarization is significant particularly at low frequencies (\citealt{Burn66, ars2011}). In the present study, we attempt to perform broadband observational visualization of a spiral galaxy, using the data of a global 3D MHD simulation of a galactic gas disk \citep{mac2013}. It aims to clarify the relation between projected observables and actual 3D structure of magnetic fields in our simulation. We introduce the numerical data in Section 2 and the method of calculation in Sections 3. The numerical results are shown in Section 4, followed by discussion and summary in Sections 5 and 6, respectively.

\section{MHD Simulation of a Spiral Galaxy}

We analyze the results of global MHD simulations of a galactic gaseous disk. The simulation was performed in the cylindrical coordinate system $(r, \phi, z')$. 
The simulation box size is defined in the radial direction as $r~({\rm kpc}) < 56$, in the azimuthal direction as $0 \leq \phi \leq 2 \pi$, and in the vertical direction as $-10 < z'~({\rm kpc}) < 10$, and is gridded evenly with cells of $(250, 256, 640)$ in the $(r, \phi, z')$ directions, respectively.  In the main domain where we address in this paper, $0 < r~({\rm kpc}) < 6$ and $-2 < z'~({\rm kpc}) < 2$, the spatial resolution of the simulation is $\Delta r=50$~pc and $\Delta z'=10$~pc. With $\Delta \phi = 2 \pi /256$, the resolution of the azimuthal direction, $r \Delta \phi$, depends on $r$ and is $25$~pc at $r=1$~kpc.

The initial condition of the simulation was an equilibrium torus threaded by weak azimuthal magnetic fields embedded in a non-rotating hot corona under the Miyamoto and Nagai potential \citep{miy1975}.  
The self-gravity of the gaseous disk, radiative cooling, and spiral-arm potential of the stellar disk were all ignored in their paper. 
%
The initial torus with the temperature $10^4~{\rm K}$ and the density $1~{\rm cm^{-3}}$ at $r=10~{\rm kpc}$ falls into the central region and forms a galactic gaseous disk. 
The resultant disk has the equatorial density of $0.1~{\rm cm^{-3}}$ at $r=2\sim3~{\rm kpc}$.  
Angular momentum of the gas in the disk is transported by the MRI, and the gas temperature increases to $10^5~{\rm K}$ due to the adiabatic compression and the Joule heating by magnetic reconnection in the disk and the boundary layer between the disk and the halo. 
Magnetic field in the disk is amplified in such a high temperature plasma. 
The plasma $\beta$ (the ratio of the gas pressure to the magnetic pressure) in the disk ranges from 5 to 100 and the averaged value becomes about 20. Note that the temperature varies with time but its fluctuation is small, because we did not consider gas cooling mechanism by radiation.

Although the disk magnetic field is tangled and becomes turbulent, structures of the global magnetic field in the disk and the halo have the compositions of the higher-order mode and ASS-like topologies, respectively. Such a duality seems to be consistent with the result of Faraday tomography for M51 \citep{fle2011}.

Both the mean and turbulent components show similar peak strengths of about 10~$\mu {\rm G}$. 
But the volume filling factor of such high strength regions is less than 0.1 in the disk and the average strength in the disk is about a few $\mu {\rm G}$. 
The magnetic energy of the turbulent component is about three times larger than that of the mean component inside the gas disk, and they become comparable each other in the halo region.

Our model exhibits a few magnetic spiral arms on the contour of the magnetic energy. But the magnetic spiral arms in the numerical results do not corresponds to the grand-design spiral such as M51, because we do not consider the cooling effect of the gas. The spiral galaxies showing grand-design spiralls (e.g., M51 and so on) and higher starforming activities are beyond the scope of this paper. See Appendix A. and \citet{mac2013} for more details of the simulation data. 
Meanwhile, IC342 is one of the targets of our model (see recent  review by \citealt{bec2016}).  
We attempt to compare IC342 with our simulation in section 5.

\section{Observational Visualization}

We consider observations of an external galaxy and calculate radio synchrotron observables such as the Stokes parameters and Faraday rotation measure. 
Since numerical simulation data has the physical values such as magnetic field vector and density at each grid point, we directly calculate the Faraday depth (FD) which is a integral of the LOS component of magnetic field weighted with the density. 
If the Faraday rotation occurs at the foreground of the synchrotron emitting source, the FD is equal to the Faraday rotation measure.

\begin{figure}
\begin{center}
\includegraphics[width=8cm]{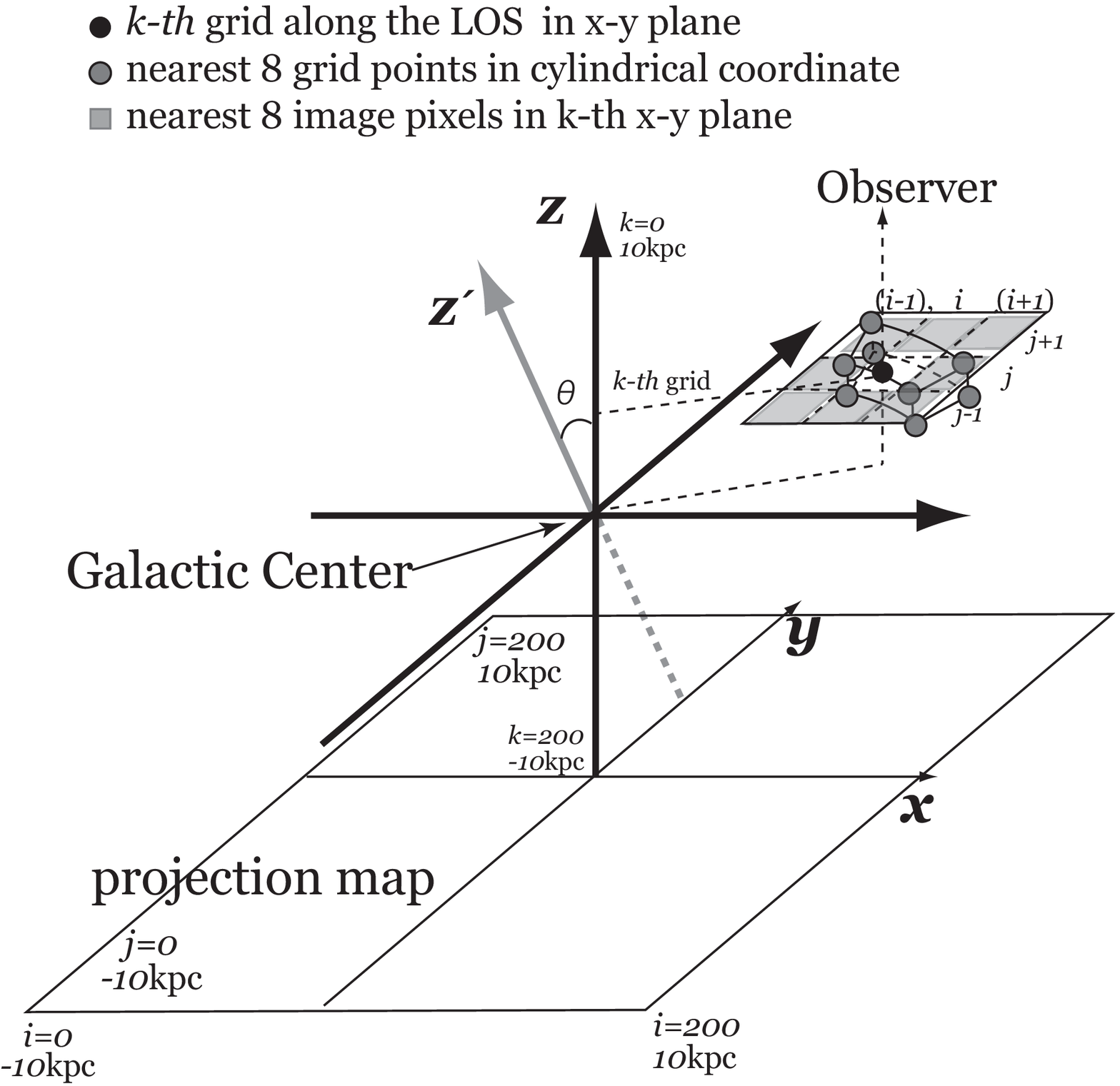}
\end{center}
\caption{
Schematic drawing of the visualization region. Black axises ($x, y, z$-axis) show the coordinate of the observational visualization. LOS components are along the $z$-axis. $z^{'}$-axis shows the coordinate which is used in the MHD simulation of the galactic disk. A sectorial box denotes the computational grid box using MHD simulations which are adopted in the cylindrical coordinate system. Gray points are the grid points of the MHD simulation and black point is a grid point along the LOS $(i, j, k)$. $\theta$ is an inclination angle of the galaxy.  Gray hatched region show nearest eight cells of the image pixel $(i, j)$ in $k$-th plane. 
}\label{fig1}
\end{figure}

Fig.~\ref{fig1} shows the schematic drawing of the visualization region.
The computational volume of the visualization is a cube $(x, y, z)$ with sides of 20~kpc, which is centered on the galactic center. Two-dimensional maps are constructed with $(N_{x}, N_{y})=(200, 200)$ image pixels centered at the projected galactic center which corresponds to the bottom square in $xy$-plane on Fig.~\ref{fig1}.   
Hence the image resolution becomes 100~pc.  
This resolution corresponds to $\sim 2"$, 
if we assume a galaxy is placed in the Virgo cluster at a distance of $\sim 20$~Mpc. 

For each image pixel, we integrate observables along the $z$-axis. We incline the $z'$-axis of the cylindrical coordinate of the data along the $y$-$z$ plane and the inclination angle, $\theta$, is defined as the angle from the $z$-axis to the $z'$-axis; $\theta=0~\degr$ and $\theta=90~\degr$  correspond to face-on and edge-on view, respectively.

\subsection{Procedure of Visualization}
\
The integration for the image pixel at $(i, j)$ is performed as follows. The local FD within a certain computational cell is given by $0.81~n_{\rm e} B_{\parallel} \Delta l$, where $\Delta l=100$~pc is the line element. The electron density, $n_{\rm e}$, and the LOS component of the magnetic field, $B_{\parallel}$, are derived from interpolation of the nearest eight cells of the cylindrical coordinate in the data (see Fig~\ref{fig1} black and gray points). The FD along the LOS is calculated by integrating local FDs. 
\begin{equation}
FD_{{\rm ave},i, j, k} =\sum_{l=1}^{k} 0.81~ n_{\rm e}  B_{\|} ~\Delta l,
\end{equation}
We also integrate the optical depth,
\begin{equation}
\tau_{i, j, k} = 8.235\times 10^{-2} \nu^{-2.1} \sum_{l=1}^{k} T^{-1.35} n_{\rm e}^2 \Delta l,
\end{equation}
so as to consider free--free absorption. Here $T$ is the electron temperature derived from interpolation of the nearest eight cells of the image pixels in the data (gray hatched region in Fig.~\ref{fig1}), and $\nu$ is the frequency.

Following \citet{sok98}, the intrinsic 
polarization degree (PD) $W$ of the polarization emitted at each computational cell is estimated from the frequency independent depolarization:
\begin{equation}
W = \frac{1+3.5 q^2}{1+4.5 q^2 + 2.5 q^4}
\end{equation}
Here, $q=\overline{b_{\bot}}/\overline{B_{\bot}}$, which corresponds to the ratio of the turbulent field to the mean field perpendicular to LOS.

Specific Stokes parameters are given by the formulae in the literature \citep{sun08, wae09}. With the cosmic-ray electron energy spectral index, $p$, and the coefficients, $g_1(p)$ and $g_2(p)$ (Appendix B), observable specific Stokes parameters of synchrotron radiation emitted at the $k$-th grid are given by
\begin{equation}
\Delta I= g_1(p) F \Delta l,
\end{equation}
\begin{equation}\label{eq:dQ}
\Delta Q=g_2(p) F \cos{(2\chi_{i, j, k})} ~ W ~  \Delta l,
\end{equation}
and
\begin{equation}\label{eq:dU}
\Delta U=g_2(p) F \sin{(2\chi_{i, j, k})} ~  W~ \Delta l,
\end{equation}
where
\begin{equation}
F = C(r) B_{\bot}^{(1+p)/2} (2 \pi\nu)^{(1-p)/2} e^{-\tau_{i, j, k}}.
\end{equation}
The term $e^{-\tau_{i, j, k}}$ gives free--free absorption of synchrotron emission.  

Faraday rotation of the PA in Eqs. (\ref{eq:dQ}) and (\ref{eq:dU}) is given by
\begin{equation}
\chi_{i, j, k} = \frac{1}{2}\cos^{-1}{\left( \frac{B_x}{|B_{\bot}|} \right)} + FD_{{\rm ave},i, j, k} \lambda^2 ,
\end{equation}
where the first term gives the intrinsic PA. We integrate $\Delta I$, $\Delta Q$, and $\Delta U$ along the LOS to obtain the specific Stokes parameters ($I$, $Q$, and $U$). 
Because we do not consider the distance from the observer to the galaxy, absolute values of Stokes parameters are rather arbitrary and only relative brightness is meaningful in this paper. It means that the length of the LOS is about 20~kpc which corresponds to the diameter of the galaxy.

We also present the PA,
\begin{equation}\label{eq:chi}
\chi = \frac{1}{2} \tan^{-1} \left(\frac{U}{Q} \right), 
\end{equation}
and the PD, 
\begin{equation}
\Psi = \frac{P}{I}=\frac{\sqrt{Q^2+U^2}}{I},
\end{equation}
where we do not consider circular polarization, which is weak in the context of this study.

According to \citet{wer2015}, electrons with energies greater than 100~GeV are confined to their source regions and trace the cosmic-ray source distribution closely. Because cosmic-ray electrons responsible for radio emission in 8 GHz in a magnetic field of 10~$\mu {\rm G}$ have energies of 10~GeV, the cosmic-ray electron is diffused from the magnetic spiral arm. The diffusion lengths of GeV electrons is about 1kpc during their lifetime of a few $10^7$ years. Since this time scale is much more than the resolution in our model, the assumption of local equipartition is not valid. However, the diffusion of cosmic-ray is beyond the scope of this paper. 
If the total energy of cosmic-ray is comparable to the thermal energy and energy equipartition with magnetic fields is assumed, this contradicts $\beta=10$ as derived in the MHD model (Appendix A). 
The energy spectral index $p=3$ is adopted as a typical value \citep{sun08}.

\section{Results}

We present the result of the simulation at a few tens of rotation periods at $r =10~{\rm kpc}$ as a representative case. It corresponds to about 48 billion years after the start of the simulation, at which both regular and turbulent magnetic fields are well-established by the MRI-Parker instability. 
Below, we show the results of face-on view, edge-on view, and intermediate inclination view in this order.

\subsection{Face-on View} 

\begin{figure*}
\begin{center}
\includegraphics[width=16cm]{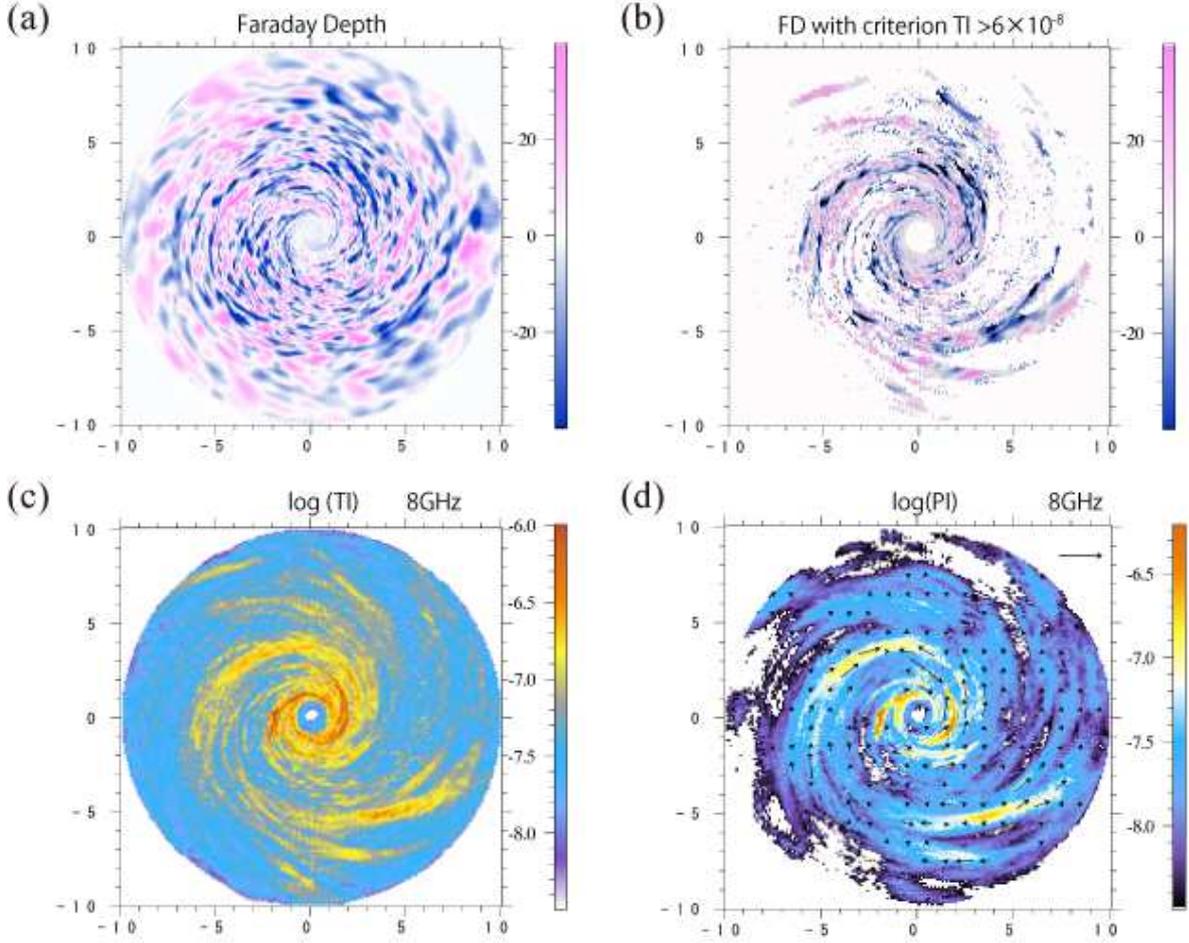} 
\end{center}
\caption{
(a) FD contour map of the simulated spiral galaxy for an inclination angle $\theta=5 \degr$. Both horizontal and vertical axes have units of kpc, and the FD has units of rad~m$^{-2}$. (b) Same as (a) but only the pixels with TI(8~GHz) larger than the average of TI(8~GHz) within the image are shown. 
(c) Simulated synchrotron total intensity at the observed frequency of 8~GHz. (d) Simulated polarized intensity at the observed frequency of 8~GHz, all in units of Jy/sr. 
Arrows indicate the magnetic vector $(2\chi+90\degr)$ whose strength is normalized by the maximum strength in the map. The sample vector indicates the strength of magnetic field which is 20\% of the maximum one. 
}\label{fig2}
\end{figure*}

Fig.~\ref{fig2}a shows the FD ($FD_{\rm ave}$) map for the case of inclination angle $5 \degr$. It is clearly seen that the FD distribution is patchy and complex due to turbulent magnetic fields. 
Fig. \ref{fig2}b displays FDs in above-average bright regions of the simulated synchrotron total intensity (TI) at 8~GHz in order to highlight FDs at brighter regions.

TI and polarized intensity (PI) at the observing frequency of 8~GHz are shown in Fig.~\ref{fig2}c and \ref{fig2}d, respectively. 
Arrows show the magnetic vector, $2\chi+90 \degr$, with $\chi$ obtained from the Stokes Q and U (Eq.\ref{eq:chi}), whose strength is normalized by the strength of he maximum one in the map. 
The sample vector on Fig.~\ref{fig2}d denote 20\% of the strongest magnetic strength.  
At some places, the direction of the magnetic field turns $\sim 180$ degrees across an arm/interarm. This is because the magnetic spiral arms have neutral current sheets.

TI clearly shows spiral structure. Such patterns are also seen in PI, where PI and TI are similar to each other. 
Actually, Faraday rotation is weak at 8~GHz, and the magnetic vector, which is parallel to the spiral pattern of PI, traces the direction of magnetic fields. This indicates that TI and PI reflect the spiral magnetic-field structure inside the gas disk rather than magnetic fields in the halo.

\subsection{Edge-on View}

\begin{figure*}
\begin{center}
 \includegraphics[width=15cm]{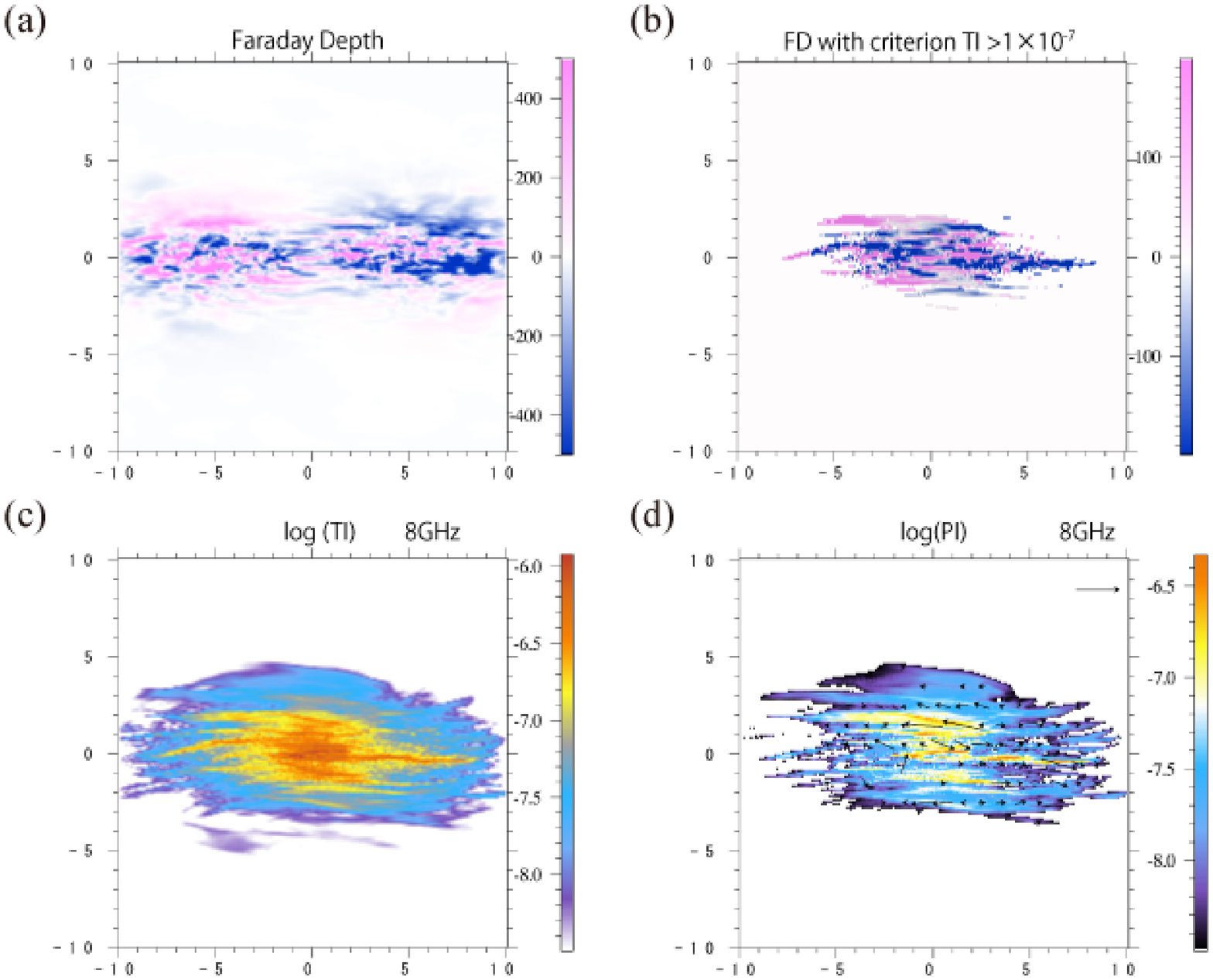}
\end{center}
\caption{
Same as Fig.~\ref{fig2} but for $\theta=85 \degr$.
}\label{fig3}
\end{figure*}

Fig.~\ref{fig3}a shows the FD map for $\theta=85 \degr$. The FD is mainly derived from the gas at the mid-plane of the LOS, because the azimuthal magnetic field is dominant and only the LOS component of magnetic fields can contribute to the FD. Similarly, FD around the rotation axis ($x\sim 0$), at which denser gas exists, is rather small.
Because the azimuthal magnetic field is dominant in the gaseous disk, the azimuthal magnetic field becomes parallel to the LOS around the z-axis. 

There are reversals of the sign of FD along the radial direction. The reversals mean multiple inversions of the direction of the global azimuthal magnetic field. Actually, azimuthal magnetic fields, which emerged from the disk owing to the Parker instability, show periodic reversals. Such positive and negative FDs can be seen as stripes in the vicinity of the rotation axis. The length scale of emerging flux tubes from the disk to the halo are about a few~kpc, although the length depends on their radius. The striped FD patterns could be unique evidence that the magnetic flux flows out into the galactic halo 
successfully. Note that the pattern seems simply random around the rotation axis, if we focus on brighter regions (Fig.~\ref{fig3}b). This means that it is difficult to identify the global patterns, if we only have low-sensitivity polarization data.

Figs.~\ref{fig3}c and \ref{fig3}d show the maps of TI and PI at 8~GHz, respectively. Overall, TI becomes larger towards the galactic mid-plane, because the global magnetic field is strongest in the disk. Structures seen in PI are similar to those seen in TI, and the magnetic vector nicely traces the global azimuthal magnetic field. 

\subsection{Intermediate Inclination Angles}

\begin{figure*}
\begin{center}
\includegraphics[width=17.5cm]{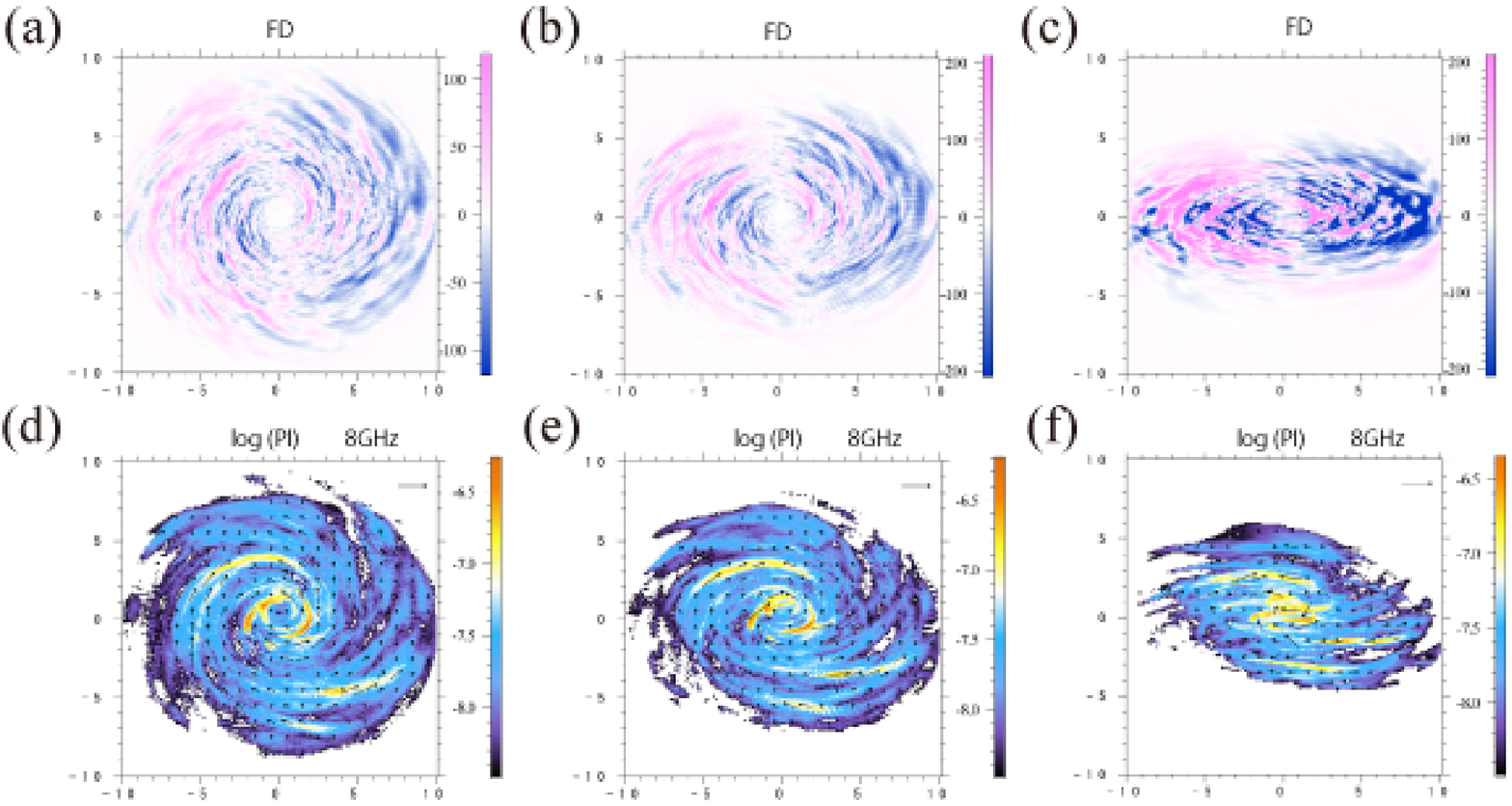}
\end{center}
\caption{
Maps of the simulated observables. Top and bottom panels show the FD and PI at 8~GHz. Also shown are the magnetic vector angles $(2\chi+90\degr)$ as arrows. Panels from left to right show the results for the inclination angles $\theta=25\degr, 45\degr, 70\degr$, respectively. Both horizontal and vertical axes have units of kpc. 
}\label{fig4}
\end{figure*}

Fig.~\ref{fig4} shows the results for the inclination angles $\theta = 25\degr$, $45\degr$, and $70\degr$ from 
left to right, respectively. As already mentioned in Fig.~\ref{fig3}, the FD is relatively small along the rotation axis of the galaxy ($x \sim 0$), since the dominant, azimuthal magnetic field is not parallel to the LOS and hence does not contribute to the FD. In contrast, PI, which intrinsically depends on the magnetic fields perpendicular to the LOS, is maximum around $x\sim 0$ and weaker away from $x\sim 0$.

We find that the topology of the magnetic vector field in PI is different between small and large viewing angles, although we have used the same numerical data of a spiral galaxy. 
The magnetic vector field apparently can be explained with 
a mixed mode of ASS and higher mode for $\theta = 25\degr$. 
As we increase $\theta$, a ring-like structure partly mixes in the pattern.

\begin{figure}
\begin{center}
\includegraphics[width=8cm]{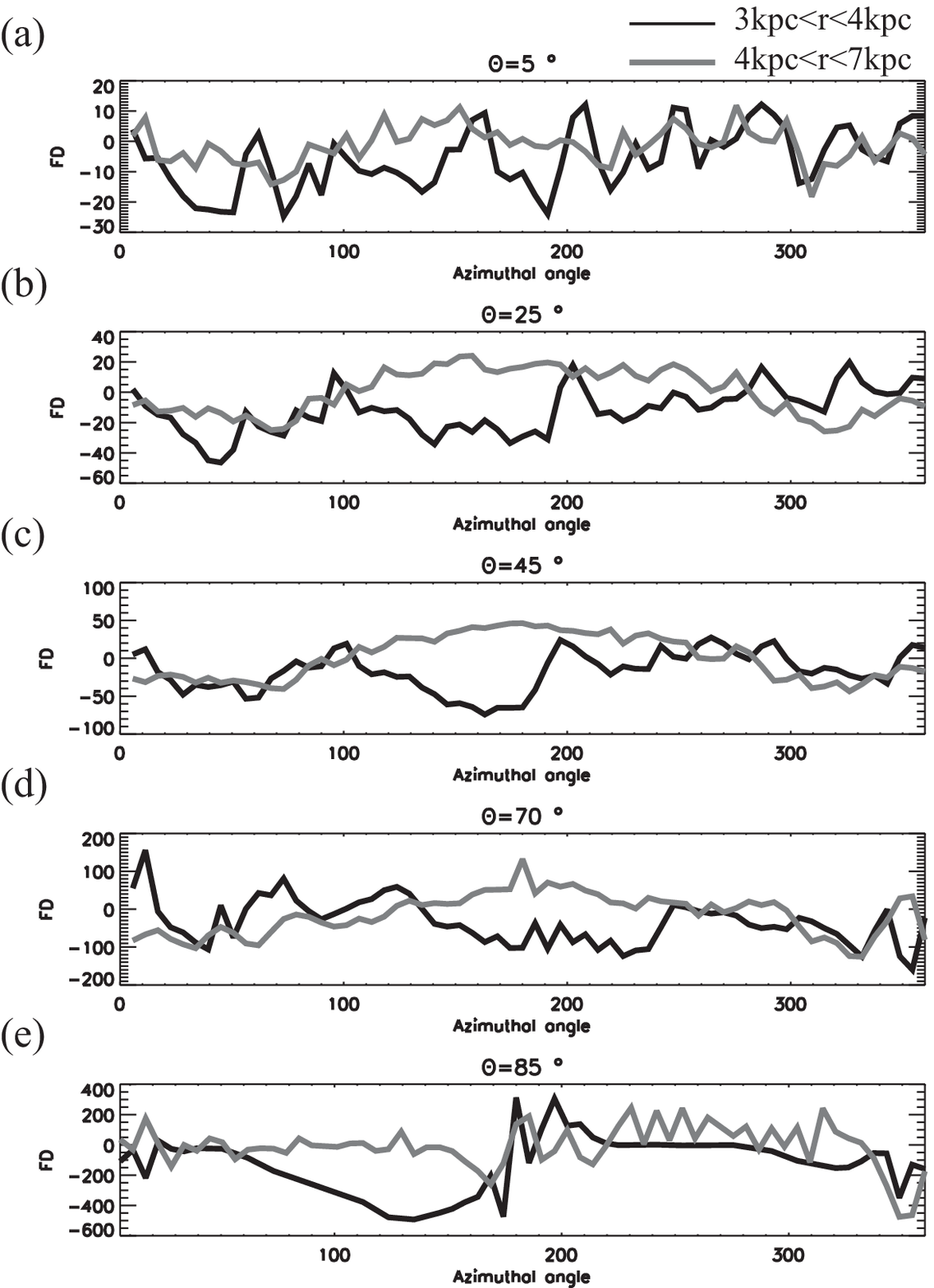}
\end{center}
\caption{
The azimuthal distribution of the FD. The black curves show the FD in the region of $3 < r [{\rm kpc}] < 4$, and 
gray curves show that in the region $4 < r [{\rm kpc}] < 7$. 
Top and bottom panels show the results of $\theta = 5\degr$ and $85\degr$, respectively. 
}\label{fig5}
\end{figure}

Fig.~\ref{fig5} shows an azimuthal variation of FD, which was calculated by integrating value of the FD in ranges of $3 < r [{\rm kpc}] < 4$ and 
$4 < r [{\rm kpc}] <7$, each of which are indicated with black and gray curves, respectively. 
The dependence of the inclination angles is shown from top to bottom.  
The magnetic reversals found in the range of $ < r [{\rm kpc}] < 4$, as indicated by a changes in sign of FD, is attributed to integration range. 
The number of the reversals is comparable to the turbulent scale.
In the face-on view, these structures indicate that the reversal of the vertical components along the LOS are observed. 
When the inclination angle becomes large, there are less reversals in the FD. 
Essentially, the higher mode is expected in our numerical simulation data because there are a few magnetic spiral arms. 
On the other hands, since the integrated region of the gray curves are larger than the scale length of the turbulence. 
The fluctuations in the curves are disappeared and $\phi$-FD relations show a sinusoidal distribution (\ref{fig5}b, c, d), 
which are typical structure of ASS. 
The FD curves of the outer region (gray one) still show the reversals of FD, which is considered to comborution of the ASS and the higher mode.
%

\section{Discussion}

\subsection{LOS Distribution of Magnetic Fields}

\begin{figure}
\begin{center}
\includegraphics[width=8cm]{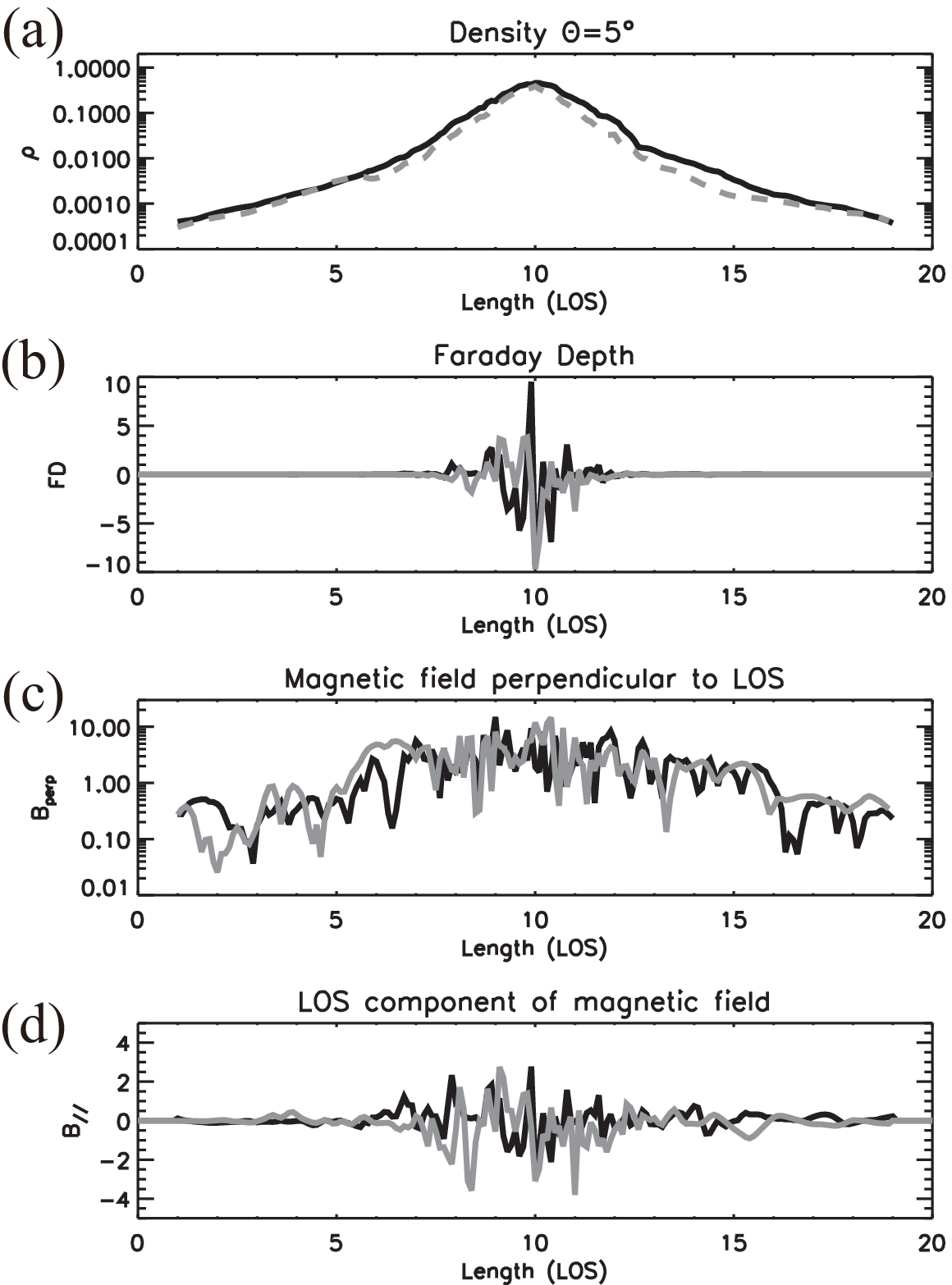} 
\end{center}
\caption{
Distribution of physical quantities along the LOS (in kpc) for the inclination angle $\theta = 5\degr$. Panels show (a) the gas number density in cm$^{-3}$, (b) the local FD (in rad~m$^{-2}$) within each computational cell with a depth of 100~pc (c) $B_{\bot}^2$ in $({\rm \mu G})^2$, and (d) $B_{\|}$ in $\mu$G. Black and gray curves show the values toward  the inter-arm (-3~kpc, 0~kpc) and the arm (2~kpc, -0.5~kpc), respectively. The galactic midplane is located at $\sim 10$~kpc.
}\label{fig6}
\end{figure}

To link the projected observables to actual 3D structures, we plot the physical quantities along the LOS. Fig.~\ref{fig6} shows the cross sections toward the outer inter-arm(black) and the inner arm (gray) in the case of $\theta = 5 \degr$. As seen, the density distribution is smooth; the average number density is about 0.03~cm$^{-3}$ in the disk and about $3 \times 10^{-3}$~cm$^{-3}$ in the halo. On the other hand, the LOS component of the magnetic field is disturbed and the local FDs within each computational cell indicate a complex distribution. The integration thus behaves like a random walk process; the integrated FD toward the arm is $-18$~rad~m$^{-2}$ although the maximum local positive FD along the LOS reaches about $4$~rad~m$^{-2}$.

Note that $B_{\|}$ has some peaks around LOS $\sim 8$~kpc ($\sim 2$~kpc far from the equatorial plane). The peak magnetic fields are formed by the magnetic wind resulting from the flotation of the Parker instability. However, because the halo density is low, the FD induced by such magnetic fields does not significantly contribute to the total FD. 

\begin{figure}
\begin{center}
\includegraphics[width=8cm]{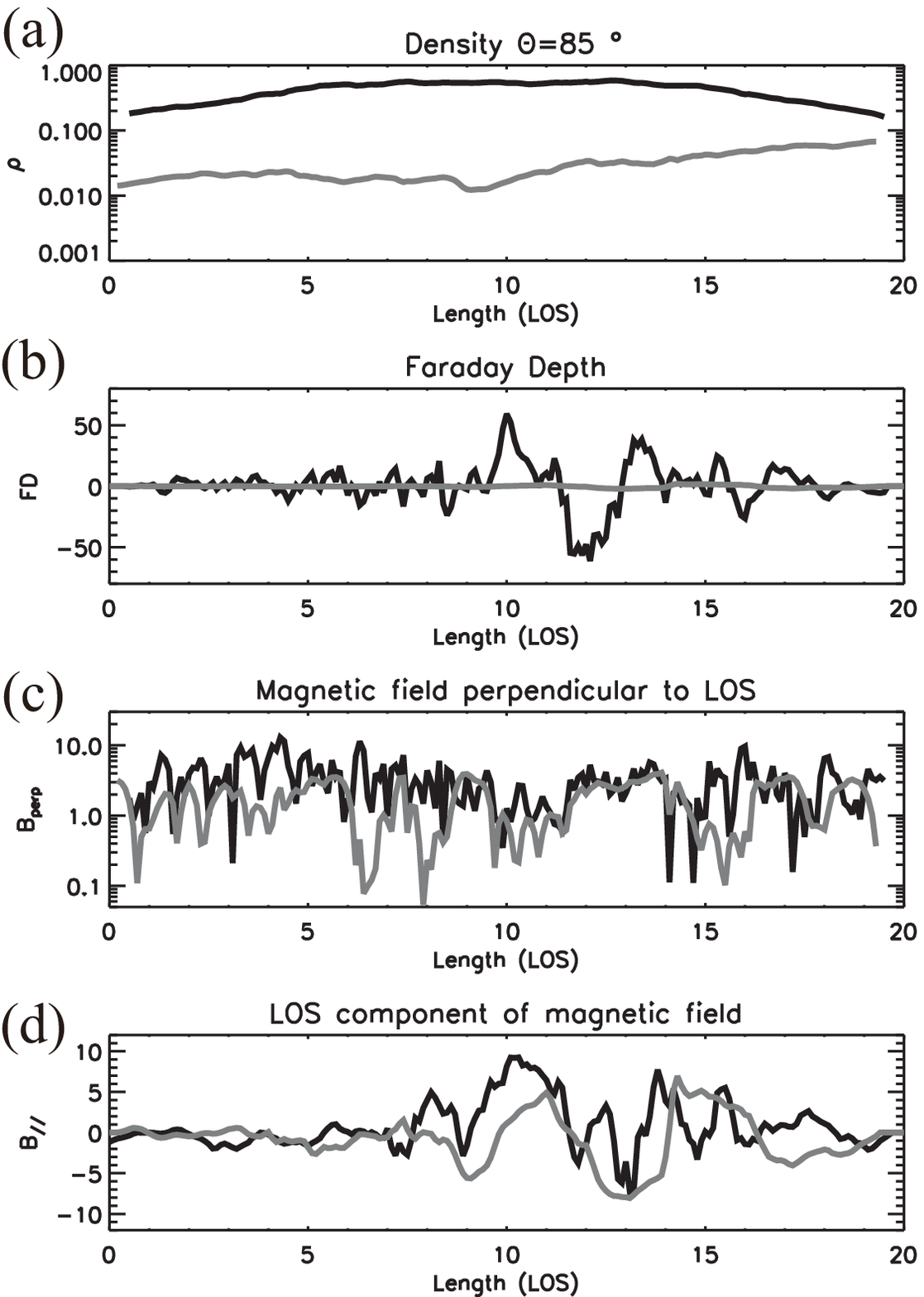}
\end{center}
\caption{
Same as Fig.~\ref{fig6} but for $\theta =85\degr$. Black and gray curves show the values toward the disk (3~kpc, 0~kpc) and the halo (-3~kpc, 3~kpc), respectively.
}\label{fig7}
\end{figure}

The same plots as Fig.~\ref{fig6} but toward the disk (black) and halo (gray) in the case of $\theta =85\degr$ are shown in Fig.~\ref{fig7}. The distribution of the local FD shows complex structure reflecting multiple reversals of $B_{\|}$ due to the inversion of the azimuthal magnetic field in the radial direction. 
$B_{\|}$ and $B_{\perp}$ have comparable strengths of $\sim 10$~$\mu$G, but FDs are very different and the FD in the halo is very small due to the low density in the halo.

The resultant cumulative FD toward the disk is 151~rad~m$^{-2}$, although the local FDs show about $-60~{\rm rad~m^{-2}}$.

\subsection{Is X-field Missing due to Reduced Outflows?}

Our visualizations for the edge-on view do not indicate the X-shape structure of the magnetic vector, which is seen in e.g. NGC253 \citep{hee2009}. Even though we choose a higher frequency such as 8~GHz, at which both Faraday rotation and depolarization are negligible and the magnetic vector thus traces the intrinsic direction of magnetic fields, the magnetic vectors indicate not vertical but azimuthal magnetic fields (e.g. Fig.\ref{fig3}b). 

In our global MHD simulation of a spiral galaxy, it is possible that the growth of vertical magnetic fields is reduced on some level and thus the X-field does not exceed the azimuthal magnetic field. This is because the outflow velocity was $\sim 30~{\rm km/s}$, which is $\sim 10$~\% of that observed in starburst galaxies. A possible reason for such a low outflow velocity is that we ignored cosmic-ray pressure and/or we assumed high gas pressure initially. The effects of cosmic rays on magnetic instability were studied by several authors (e.g., \citealt{kuw2004, kha2012, suz2014, kuw2015}). According to these studies, the cosmic-ray pressure can support the growth of a magnetic instability such as the MRI or Parker instability, although the results depend on the diffusion of cosmic-rays and magnetic field structure. When starburst becomes dominant, the stellar wind could also produce powerful winds and cosmic-ray pressure could assist the acceleration of galactic winds.

\subsection{Comparison with IC342}


The FD reversal within the spiral arms are found in the galaxy IC342 which shows a low star formation rate and unclear spiral arms \citep{bec2015}. 
To compare to the observations, we made a plot of the FD in the spiral arms (see Fig.~\ref{fig2}b). The periodical reversals of FD are observed in the spiral arms. Because the scale length correlates with the disk scale height, the reversal length of the FD signs are proportional to the radius. It means that these reversals are caused by the Parker instability.  These results support to the observational results for IC342 and theoretical predictions of the Parker instability.  
Fig.~\ref{fig5}b shows the relation between azimuthal angle and FD inside $4~{\rm kpc} < r < 7~{\rm kpc} $ for $\theta = 25\degr$, whose inclination angle seems to be IC342. The tendency of azimuthal angle - FD relation shows similar with IC342. 

Since we ignored the cool components of galactic gas disk, the gas density and magnetic field strength will be underestimated. When we will consider the multi-temperature distribution of galactic disk, these discrepancy will be resolved.

\subsection{Enhancement of global magnetic fields}

In our simulation, the energy of turbulent magnetic fields is three times larger than that of the mean magentic field which is dominated by the azimuthal component in the disk. 
The peak strength of the mean azimuthal component, which corresponds to the magnetic spiral arm, is 70\% of that of the turbulent component. 
These magnetic spiral arms are much fainter than the grand-design spiral of the galaxies. 
This is because the temperature inside the disk becomes about $10^5$~K, the turbulent cell becomes large. According to the \citet{mac2006} which reported about the state transition of X-ray binaries, when the cooling effect worked in the accretion disk, the accretion disk shrinked  in vertical direction with magnetic flux and mean magnetic fields enhanced inside the accretion disk. For the accretion disk case, the accretion disk became magnetically supported when the shrinking was stopped and the disk scale height of magnetically supported disk became 20\% of the gas pressure (high temperature) disk. The turbulent components of magnetic fields are dissipated inside the thin disk and the ordered field remained inside the disk.   
We speculate that magnetic field evolution through the gas heating and cooling qualitatively work in a similar manner between the galactic disk and the accretion disk. First, a part of nutral gas is ionized by supernovae, OB-star UV radiation and magnetic reconnection, etc. These ionization mechanisms are strongly couples with magnetic fields. Second, turbulence is generated and magnetic field is enhanced by the MRI and Parker instability in the radial direction. Third, the ionized gas returns to neutral if the density is high enough to work the molecular cooling. The magnetic flux shrinks in vertical direction and the mean magnetic field enhances inside the disk. 
However, the scale height of warm HI components of galactic gas disk is comparable to the high temperature plasma, since there are extra energy inputs caused by the super novae and magnetic activities. Therefore,  in the case of the galactic gaseous disk, we expect that the turbulent field is not much dissipated and the ordered magnetic energy will be comparable to the turbulent magnetic energy. In addition, the spiral potential of the galactic star disk has an important effect to produce the ordered magnetic fields.

\section{Summary}

We summarize the points discussed in the present work:

\begin{itemize}

\item Polarized intensity (PI) becomes stronger towards the rotation axis of the galaxy, while Faraday depth (FD) shows the opposite tendency. This is because the azimuthal component of magnetic field is predominant inside the disk and thus the LOS component contributing to FD and the perpendicular component inducing PI depend on the distance from the rotation axis. These tendencies do not depend on the inclination angle of the galaxy.

\item Total intensity (TI) and PI(8\,GHz) are both stronger along denser magnetic spiral arms, where depolarization is insignificant at such high frequencies. The polarization angle (PA) traces the orientation of the azimuthal magnetic field inside the disk. 

\item An ASS-like structure appears in the magnetic-vector map for the case of $\theta=85\degr$, and one sine curve is seen in the azimuthal angle -- FD relation. Meanwhile, since a few magnetic spiral arms are present in the numerical simulation data, a sine curve with three to four periods is expected in the azimuthal angle -- FD relation in the case of $\theta=5\degr$. The topology, however, cannot be clearly classified as BSS with an inclination angle less than $70 \degr$. 

\end{itemize}

An X-shape vertical structure is not observed in our results. It may be ascribed to the fact that our numerical results have weak outflows produced by Parker instability. If we include the effect of cosmic rays, the growth rate of the Parker instability becomes large and the outflow speed becomes high enough to explain the observations. Our assumption that the cosmic ray density is proportional to the magnetic energy should be improved in future works, so as to examine the structure in detail.

Although we ignored the spiral potential of a stellar disk, the spiral potential will be important in forming a magnetic spiral structure such as the BSS configuration. We plan to add this effect in the near feature. \citet{suz2015} performed a 3D isothermal simulation of a galactic center region with axisymmetric gravitational potential, and they found that spatially dependent amplification of the magnetic field possibly explains the observed circular motion of gas in the galactic center region. Simulations of the distribution of cosmic rays are calculated by \citet{wer2015}. They showed that electrons with energies greater than 100~GeV are confined to their source regions and trace the cosmic-ray source distribution closely. And lower energy electrons diffuse to the inter-arm region. 

Finally, we did not employ Faraday tomography, which is clearly complementary to our work and may be more powerful so as to deproject the LOS structure. It is, however, not trivial so far how we can translate the Faraday spectrum into real 3D structure of the galaxy \citep{ide2014}. We need further studies of both observational visualization and Faraday tomography for numerical data.

\section*{Acknowledgements}
We are grateful to Dr. R. Matsumoto, Dr. K. Takahashi, and Dr. S. Ideguchi for useful discussion. Numerical computations were carried out on SX-9 and XC30 at the Center for Computational Astrophysics, CfCA of NAOJ (P.I. MM).  A part of this research used computational resources of the HPCI system provided by (FX10 of Kyushu University) through the HPCI System Research Project (Project ID:hp140170).This work is financially supported in part by a Grant-in-Aid for  Scientific Research (KAKENHI) from JSPS (P.I. MM:23740153, 16H03954,  TA:15K17614, 15H03639).


\appendix
\section{Numerical Simulation of Galactic Gas Disk} 

\citet{mac2013} performed a 3D MHD simulation of the galactic gas disk. Below, we briefly summarize the evolution of magnetic field in the simulation.

The normalization of the velocity is $v_0=(G M_0/r_0)^{0.5}$ where we assume the unit length is $r_0=1~{\rm kpc}$ and $M_0=10^{10}M_{\odot}$. 
The unit temperature is calculated by the unit velocity. 
This means that the unit velocity corresponds to the sound velocity. 
The thermal energy of the initial torus is balanced the gravitational energy which is made by the central black hole. 
They assumed the ratio of thermal energy to the gravitational energy is 0.05. 
Because the initial magnetic energy is defined by the plasma $\beta$ which is the ratio of the gas pressure to magnetic pressure, 
magnetic energy is also proportional to the unit density of gas and has been defined by  $B_0=\sqrt{\rho_0 v_0^2}$. 
In \citet{mac2013},  since the unit velocity and the unit density have been assumed about $207 {\rm km/s}$ and $1 {\rm cm^{-3}}$,  
the unit strength of the magnetic fields is about $26 {\rm \mu G}$.  
Because the initial magnetic fields have been defined at $10 {\rm kpc}$ and the ratio of thermal energy to gravitational energy at that radius have 
been assumed at $0.05$, the initial fields strength have been about $0.3 {\rm \mu G}$.

Initial weak azimuthal magnetic fields are amplified initially by the differential rotation and by the MRI. After several rotation periods, averaged plasma $\beta$ inside the disk becomes 10 which means that magnetic pressure is about 10~\% of the gas pressure. Here, the rotation period at 1~kpc corresponds to the unit time of the simulation $t_0 = 4.8 \times 10^6 {\rm yr}$.

When the disk reaches quasi-steady state, gaseous disk becomes turbulent and a strong magnetic flux forms locally, and he magnetic field strength reaches a few $\mu~{\rm G}$. The volume filling factor for low--$\beta$ ($\beta \sim 5$) is 0.1. The density distribution of the gaseous disk near the galactic center shows a weak spiral structure and the magnetic fields also appear spiral, although we ignored the spiral potential produced by the stellar disk potential.

When the averaged magnetic pressure becomes about 10~\% of the gas pressure in the disk, local magnetic pressure near the surface exceeds 20~\% of the gas pressure and the Parker instability creates magnetic loops near the disk surface. Because the magnetic field strength parallel to the seed field grows faster than that of the opposite field, parallel fields satisfy the condition of the Parker instability earlier and buoyantly escape from the disk. The magnetic flux thus floats out from the disk because of the Parker instability.

Subsequently, since anti-parallel to the seed fields remain inside the disk and are amplified by the MRI, mean azimuthal magnetic fields inside the disk reverse their direction. The reversed fields become a seed field for the next cycle. The direction of the azimuthal magnetic field in both radial and vertical directions inverts quasi-periodically. The time scale of the flotation cycle is about 10 times of a rotational period. 

This mechanism has, however, not yet been proven by observations; it is still difficult to measure global magnetic fields in galaxies even with current large observational facilities such as the Jansky Very Large Array and Atacama Large Millimeter/Submillimeter Array.

The simulation adopted the initial weak azimuthal magnetic fields. Even if only an magnetic field is assumed at the beginning, numerical results show that multiple magnetic field arms are dominant in the disk owing to the development of magnetic turbulence. Meanwhile, in the halo region where an ASS magnetic field structure is dominant, magnetic lines of force are wound loosely. 

We also have checked the dependence of the above results on azimuthal resolution, because the azimuthal resolution is sensitive to the saturation levels of MRI. The resultant ratio of the radial magnetic energy to the azimuthal magnetic energy were 0.12 and 0.13 with the azimuthal resolutions of 128 grids and 256 grids (the one that adopted in this paper), respectively. This value can be used to judge whether the most unstable wavelength of the MRI was resolved or not. From the comparison, we concluded that 128-grid model did not significantly differ from the results with 256-grid model in the nonlinear stage. The azimuthal spatial resolution adopted in our work is thus sufficient to reproduce the MRI.

\section{Synchrotron Radiation and Faraday Rotation}

When relativistic charged particles such as cosmic-ray electrons move in a magnetized plasma, their tracks follow a helical curve along a magnetic field line, emitting polarized synchrotron radiation. Considering an isotropic cosmic-ray electron distribution with the cosmic-ray electron density $C(r)$, the energy spectral index $p$, and the number density $N(\gamma) d \gamma = C \gamma^{-p} d \gamma$ between the electron Lorentz factor $\gamma$ and $\gamma + d \gamma$, specific Stokes parameters of synchrotron radiation from the electron population can be written as 
\begin{equation}
I = \int G_1(p) \omega^{(1-p)/2}B_{\perp}(r)^{(1+p)/2}C(r)dr,
\end{equation}
\begin{equation}
P=Q+iU=\int G_2(p) \omega^{(1-p)/2}B_{\bot}(r)^{(1+p)/2}C(r) e^{2i \chi (r)}dr
\end{equation}
(\citealt{sun08, wae09}), where $B_{\bot}$ is the magnetic field perpendicular to the LOS, $\chi$ is the initial polarization angle at $r$, and $\omega$ is the angular frequency. Other physical coefficients are $G_1(p)=2g_1(p) j(p)$, $G_2(p)=2g_2(p) j(p)$, 
\begin{equation}
j(p) = \frac{1}{4 \pi} \frac{\sqrt{3} q^3}{8 \pi m c^2} \left(\frac{2mc}{3q} \right)^{(1-p)/2}, 
\end{equation}
\begin{equation}
g_1(p) = \frac{1}{1+p} 2^{(1+p)/2} \Gamma \left(\frac{p}{4} - \frac{1}{12} \right)\Gamma \left(\frac{p}{4} + \frac{19}{12} \right),
\end{equation}
\begin{equation}
g_2(p) = \frac{1}{p-3} 2^{(1+p)/2} \Gamma \left(\frac{p}{4} - \frac{1}{12} \right)\Gamma \left(\frac{p}{4} + \frac{7}{12} \right),
\end{equation}
where, $c$, $m$, $q$ and $\Gamma$ are the speed of light, electron mass, electric charge, and $\Gamma$ function, respectively.

FD can be derived from the free--electron density ($n_e$) of the magnetized plasma and the magnetic field parallel to the LOS ($B_{\|}$) as 
\begin{equation}
FD = \frac{q^3}{2 \pi m^2 c^4} \int_0^x n_e B_{\|} dx \sim  0.81 \int_0^x n_e B_{\|} dx \hspace{4mm} [\rm{rad ~ m^{-2}}]. 
\end{equation}

\bsp	
\label{lastpage}
\end{document}